\documentclass[aps,pra,twocolumn,superscriptaddress,nofootinbib,longbibliography]{revtex4-2}

\usepackage[utf8]{inputenc}
\usepackage[T1]{fontenc}
\usepackage{amsmath,amssymb,amsfonts}
\usepackage{graphicx}
\usepackage{hyperref}
\usepackage{xcolor}
\usepackage{braket}
\usepackage{physics}
\usepackage{amsmath}
\usepackage{amssymb}
\usepackage{amsfonts}
\usepackage{parskip}
\usepackage{comment}
\usepackage{subcaption}
\usepackage{amsmath}
\usepackage{float}
\usepackage[capitalise]{cleveref}
\usepackage{graphicx}
\usepackage{subcaption}
\usepackage[justification=raggedright,singlelinecheck=false]{caption}
\usepackage[percent]{overpic}
\usepackage[normalem]{ulem}
\usepackage{bm}

\hypersetup{
    colorlinks=true,
    linkcolor=blue,
    citecolor=blue,
    urlcolor=blue
}

\renewcommand{\emph}[1]{\textit{#1}} % @Ashwini: replaced \bf with \emph for logical markup and ability to change easily; @Ashwini: replaced bf with it for moderation MR

\newcommand{\klatt}{\ensuremath{k_{\text{latt}}}}
\newcommand{\klattbm}{\ensuremath{\bm{k_{\textbf{latt}}}}} % workaround for bold version

\begin{document}

\title{Non-resonant laser-driven narrowing of particle velocity distributions}

\author{Ashwini Vaishnav}
\affiliation{ Luxembourg Institute of Science and Technology (LIST), 5, Avenue des Hauts-Fourneaux, L-4362, Luxembourg}
\affiliation{Université du Luxembourg, 2 Av. de l’Université, Esch-sur-Alzette, L-4365, Luxembourg}

\author{Matthias Rupp}
\affiliation{ Luxembourg Institute of Science and Technology (LIST), 5, Avenue des Hauts-Fourneaux, L-4362, Luxembourg}

\author{Olivier De Castro}
\affiliation{ Luxembourg Institute of Science and Technology (LIST), 5, Avenue des Hauts-Fourneaux, L-4362, Luxembourg}

\author{Mikhail N. Shneider}
\affiliation{ Luxembourg Institute of Science and Technology (LIST), 5, Avenue des Hauts-Fourneaux, L-4362, Luxembourg}
\affiliation{Department of Mechanical and Aerospace Engineering, Princeton University, Princeton, NJ, 08544, USA}

\author{Alexandros Gerakis}
\email[Corresponding and contact author: ]{alexandros.gerakis@list.lu}
%\email{alexandros.gerakis@list.lu}
\affiliation{ Luxembourg Institute of Science and Technology (LIST), 5, Avenue des Hauts-Fourneaux, L-4362, Luxembourg}
\affiliation{Department of Aerospace Engineering, Texas A\&M University, College Station, TX, 77843, USA}

\date{\today}

\begin{abstract} 
Stark acceleration and deceleration based techniques for generating particle ensembles with low velocity spread are useful in many experimental applications. For a given velocity distribution of a particle ensemble, these techniques accelerate or decelerate a small subset of the total population, with a low velocity uncertainty. However, narrowing the original velocity distribution by accelerating or decelerating the ensemble particles near the mean velocity is fundamentally limited and not yet explored. We present a numerical study of particle dynamics using neutral cesium atoms as an example. We investigate different interaction regimes, identify key limitations, and propose an interaction regime in which optical Stark deceleration can be used to narrow the velocity distribution of a propagating ensemble about its mean velocity. These findings have potential implications for optical manipulation and control, controlled collisions, and matter interferometry.
\end{abstract}

\maketitle

\section{Introduction}

Controlling the velocity distribution of particle ensembles is crucial for applications in atomic, molecular, and other particle beam systems. Narrow velocity distributions are desirable for cooling and trapping applications, where efficient loading requires particle velocities to lie within a specific range~\cite{Meerakker2012, Metcalf, capture_val}. In controlled molecular collision experiments, a narrow velocity spread improves the resolution of collision features~\cite{Meerakker2012, collision1, collision2}. In experiments based on atom interferometry, a broad velocity distribution causes measurement instability and reduced fringe contrast~\cite{Liu_2025}. A narrow velocity distribution can improve the efficiency of applications based on matter-wave interferometry~\cite{Fiedler2024}.

Several methods have been developed to control particle motion and produce particle ensembles with narrow velocity distributions. Resonant laser cooling based methods are widely used to cool and trap atoms, ions, and molecules to near zero velocity with very narrow velocity spreads ~\cite{Metcalf, PhysRevA.20.1521, McCarron_2018, Eschner:03}. These methods exploit closed internal transitions in repeated absorption and spontaneous emission cycles, leading to a net reduction in particle velocities. However, these approaches are limited to species that have closed cycling transitions, such as alkali-metal atoms and certain molecules% \textcolor{red}{ADD A MOLECULAR EXAMPLE HERE}
~\cite{vredenbregt_2003, FitchTarbutt2021}.

Another established method, Stark deceleration, typically employs an array of electrodes with finite geometry to generate spatio-temporally variant electric fields, allowing manipulation of polar particles via dipole interactions ~\cite{Meerakker2012, Meijer2021}. It can trap and decelerate (or accelerate) a fraction of the ensemble of particles to a narrow velocity interval. However, a Stark decelerator can only effectively manipulate polar particles, requires fast-switching high-voltage across electrodes~\cite{sfbl2018}, and typically has a large physical footprint. These constraints limit the use cases for electrostatic Stark decelerators and their integration into experimental systems.

\begin{figure}
\centering
\includegraphics[width=0.45\textwidth, angle=0, origin=c]{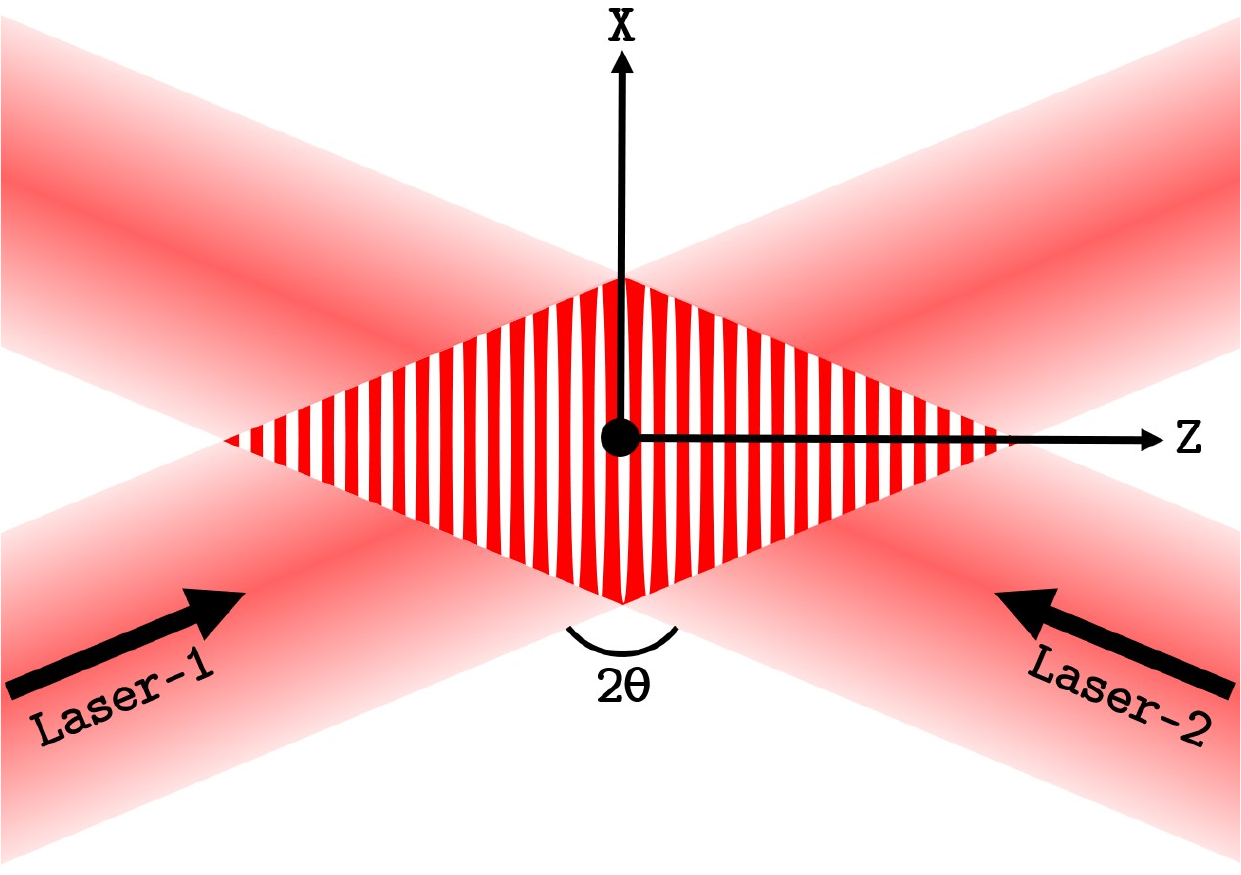}
\caption{Schematic showing formation of an optical lattice by crossing of two laser beams at crossing angle $2\theta$. The resulting periodic intensity modulation is along the $z$-axis (lattice axis). For a relative frequency difference between the laser beams, the interference pattern moves along the lattice axis.}
\label{fig:optical_lattice}
\end{figure}

\emph{Optical lattices}~\cite{GRIMM200095} provide an alternative dipole interaction-based technique that has been developed to manipulate polar and non-polar but polarizable particles through their interaction with intense laser fields~\cite{Barker2001, Barker2002}. Instead of physical electrodes, this method utilizes the interference of two mutually crossing, intense, frequency-detuned, and pulsed laser beams to generate a strong spatio-temporally varying electric field. The electric field can induce a dipole in polarizable particles; the induced dipole then interacts with the original field via dipole interaction, resulting in periodic polarizability-dependent potential variations known as an optical lattice (\cref{fig:optical_lattice}).

The velocity of the optical lattice depends on the relative frequency difference~$\Delta \omega$ and the crossing angle~2$\theta$ (see \cref{fig:optical_lattice}) between the interfering laser beams. If~$\Delta \omega$ varies over time (i.e. is chirped), the lattice accelerates or decelerates depending on the chirp slope. Such \emph{chirped optical lattices} can trap polarizable particles which follow the motion of the lattice and thus can be accelerated (or decelerated) with it~\cite{Barker2001, Barker2002}. For a given velocity distribution of polarizable particles, an optical lattice can extract a population fraction with a narrower velocity spread out of the original velocity distribution~\cite{Barker2001, Barker2002, Maher-McWilliams2012_paper, MaherMcWilliams2013_thesis, Gerakis2014_thesis}.

Despite these advances, a general method for narrowing the velocity distribution of a propagating particle ensemble about its mean velocity has not yet been explored. Resonant laser cooling mainly focuses on  facilitating trapping and confinement of particles by reducing their velocities to near zero, such as in magneto-optical trapping~\cite{Tarbutt_2015} and evaporative cooling~\cite{Davis1995, KETTERLE1996181}. Moreover, the existing Stark deceleration and chirped optical lattice methods extract only a \emph{portion of the total population} with narrow velocity spread from an initial velocity distribution via acceleration or deceleration of particles.

A working scheme for narrowing the velocity distribution of a propagating ensemble about its mean velocity requires accelerating slower particles and(/or) decelerating faster particles toward the mean velocity. Importantly, this transfer must occur while retaining the existing population in the vicinity of the mean velocity. In this study, we explore whether a chirped optical lattice can facilitate such selective manipulation, and thereby narrow the velocity and energy distribution.

We build upon earlier works~\cite{Barker2001, Barker2002, Maher-McWilliams2012_paper, MaherMcWilliams2013_thesis, Gerakis2014_thesis} on chirped optical lattices and investigate how such lattices can redistribute the particle population across the velocity space for different chirp profiles, particle distributions, and interaction approaches. We analyze the redistribution of the particle population in the velocity space and identify interaction conditions that narrow the velocity distribution. Based on this analysis, we propose and simulate an interaction scheme that can effectively narrow the velocity distribution of a propagating particle ensemble.

\section{Dipole force in an Optical Lattice}
\label{sec:DipForce}
%Dipole force%
In the presence of an external electric field, a polarizable particle acquires an induced dipole moment, which then interacts with the original electric field that created it. This gives rise to an interaction potential,
\begin{equation}
U = -\frac{1}{2} \alpha_{\text{eff}} |\mathbf{E}|^2,
\label{eq:dipole_force}
\end{equation}
where \(\alpha_{\text{eff}} \) is the effective polarizability of the particle and \textit{E} is the strength of the applied electric field. The polarizability for atoms and spherically symmetric molecules are scalar values. In general the polarizability of a molecule is a tensor. For linear molecules, however, the effective polarizability of molecular species is given by  $\alpha_{eff}=(\alpha_{\parallel}-\alpha_{\perp})\cos^2\phi + \alpha_{\perp}$~\cite{polarizibility_2006}. Here $\alpha_{\parallel}$ and $\alpha_{\parallel}$ are the parallel and perpendicular polarizability components to the molecular axis, respectively, making an angle $\phi$ with the laser polarization. Particles with a net positive polarizability feel an attractive force toward high electric field regions and are hence called \textit{high field seekers}. Similarly, when the net polarizability is negative, particles are repelled from high electric field regions and are called \textit{low field seeker species}. The particles considered here are all high field seekers. 

The high-amplitude electric fields associated with intense lasers ensure sufficiently deep dipole potentials capable of dynamic manipulation~\cite{Barker2001}. The requirements for these laser beams, namely high intensity ($10^{12} - 10^{14}$ W$\cdot$m$^{-2}$), frequency tunability (~$1$ GHz~\cite{Gerakis2014_thesis, MaherMcWilliams2013_thesis}), and far-off resonance wavelength, are fulfilled in contemporary pulsed laser systems~\cite{Bak2022, 5th_stage} used in optical lattice based experiments. For two intersecting laser beams, the net dipole potential variation is the result of the interference of both beams,
\begin{equation}
U(x,y,z,t) = -\frac{\alpha_{\text{eff}}}{2}(|\mathbf{E_1}(x, y, z, t) + \mathbf{E_2}(x, y, z, t)|^2).
\label{eq:potential_}
\end{equation}

Along the lattice axis ($z$-axis in \cref{fig:optical_lattice}), dynamics of the particles interacting with the lattice are predominately affected by the slowly varying term of ~\cref{eq:potential_}~\cite{Barker2001, MaherMcWilliams2013_thesis},
\begin{equation}
U(z,t)
=
-\frac{\alpha_{\mathrm{eff}}}{\varepsilon_0 c}
\sqrt{I_1(t)I_2(t)}
\cos\!\left[
\left|\boldsymbol{k}_2-\boldsymbol{k}_1\right|z
-
(\omega_2-\omega_1)t
\right].
\label{eq:slow varying}
\end{equation}

Here, $\varepsilon_0$ is the permittivity of free space, $c$ is the speed of light in vacuum,  $I_{1}(t)$ and $I_{2}(t)$ are the intensities of the laser beams at time~$t$. For laser beams with propagation vectors $\boldsymbol{k_1}$ and $\boldsymbol{k_2}$, the slow varying term results in steep and periodic intensity modulation along the difference vector $\klattbm=\boldsymbol{k_2 - k_1}$. $\klattbm$ is called the \emph{lattice vector} and its direction is termed the \emph{lattice axis} ($z$-axis in \cref{fig:optical_lattice}). For the laser wavelength $\lambda$ and crossing angle between the two lasers $2\theta$, $k_{\text{latt}}=4\pi \sin(\theta)/\lambda$~\cite{Maher-McWilliams2012_paper}. The difference in frequencies is called the relative frequency $\Delta \omega = \omega_2 - \omega_1$. As the relative frequency between the laser beams changes, the lattice moves and manipulates the dynamics of the ensemble particles along the lattice axis. Thus, the one-dimensional variation in dipole potential along the lattice axis ($z$) is,
\begin{equation}
 U(z, t) = -\frac{\alpha_{\text{eff}}}{\varepsilon_0 c} \sqrt{I_1(t) I_2(t)} \cos[\klatt z - \{t\Delta \omega(t)\}].
\label{eq:potential}
\end{equation}

The lattice moves with phase velocity $v_l(t) = (1/k_{\text{latt}}) \frac{d}{dt}[t\Delta \omega(t)]$ at a given time. The steep potential wells of the lattice trap particles and drag them with the lattice movement. Thus, the lattice accelerates (or decelerates) trapped particles, controlled by the lattice velocity.

%Equation of motiona and phase space things%
If the frequency difference between the laser beams is changed linearly with time (i.e. is linearly chirped) such that $\Delta \omega(t) = \Delta \omega_0 + \beta t/2$, one can write the equation of motion in the laboratory frame of reference as,
\begin{equation}
\frac{d^2z}{dt^2} = -a(t)\sin(k_{\text{latt}}z - \Delta \omega_0t - \beta t^2/2).
\label{eq:eqn_of_motion}
\end{equation}

Here, $a(t)=(\alpha_\text{eff}k_{\text{latt}}/\varepsilon_0 cm) \sqrt{I_1(t) I_2(t)}$ is the maximum acceleration possible with the lattice potential gradient for a particle having mass m,  $\Delta \omega_0$ is the initial frequency difference, and $\beta$ is the chirp rate; the sign of $\beta$ decides whether the lattice accelerates or decelerates.

To understand how the dynamics of particles evolve under the effect of the lattice force, it is useful to study the particle motion in the \emph{lattice frame}, that is, the accelerated frame moving with the lattice. After following a dimensionless transformation ({similarly to the one used in ~\cite{Barker2001}) $k_{\text{latt}}z=Z$, $\Delta \omega_0t=\Omega$, $\beta t^2/2=T^2$ and $\Theta=Z-\Omega-T^2$, the equation of motion in the lattice frame becomes
\begin{equation}
\frac{d^2\Theta}{dT^2} = -\frac{2}{\psi}\sin\Theta-2.
\label{eq:eqn_of_motion_phase_space}
\end{equation}

The dimensionless parameter $\psi=\beta/a k_{\text{latt}}$ is the ratio of lattice acceleration $\beta/k_{\text{latt}}$ to the maximum acceleration $a$ caused by the potential gradient. The motion of a particle in the transformed frame is described by the phase coordinate $\Theta$, and the phase velocity $\eta=d\Theta/dT$. The potential variation in the lattice frame is stationary and follows $U(\Theta)=-\int (m/k_{\text{latt}}^2)(d^2\Theta/dt^2)\,d\Theta$ ~\cite{Barker2001, Gerakis2014_thesis}, yielding,
\begin{equation}
U(\Theta) = \frac{m}{k_{\text{latt}}^2}\left( \Theta - \frac{\cos\Theta}{\psi} \right).
\label{eq:potential_on_phase}
\end{equation}

The stable equilibrium points occur at $\Theta=2n\pi-\sin^{-1}\psi$ and the saddle points at $\Theta=(2n-1)\pi+\sin^{-1}\psi$. The depth of the potential well ($\Delta U$) is the difference in the height of the potential at the saddle point and the nearest equilibrium point (see \cref{fig:phase_space}). The maximum phase velocity spread allowed for trapping is given by $2 \sqrt{2\Delta U/m}$~\cite{Barker2001}.

The particle velocity at a given time can be transformed back to the laboratory frame as
\begin{equation}
v = \frac{\Delta\omega_0 + \beta t}{k_{\text{latt}}}+\sqrt{\frac{\beta}{2}} \frac{\eta}{k_{\text{latt}}}.
\label{eq:velocity_lab_frame}
\end{equation}

The term $(\Delta \omega_0 + \beta t)/k_{\text{latt}}$  corresponds to the lattice velocity while the variation $\sqrt{\frac{\beta}{2}} \frac{\eta}{k_{\text{latt}}}$ is due to the lattice force. For a trapped particle, $\eta$ always remains bound in the vicinity of zero, ensuring that the particle moves with the lattice. For an untrapped particle, $\eta$ and the velocity variation term diverge, suggesting that the particle velocity keeps drifting away from the lattice velocity.

\begin{figure}[h]
    \centering

        \begin{subfigure}{1\linewidth}
        \centering
        \begin{overpic}[width=\linewidth]{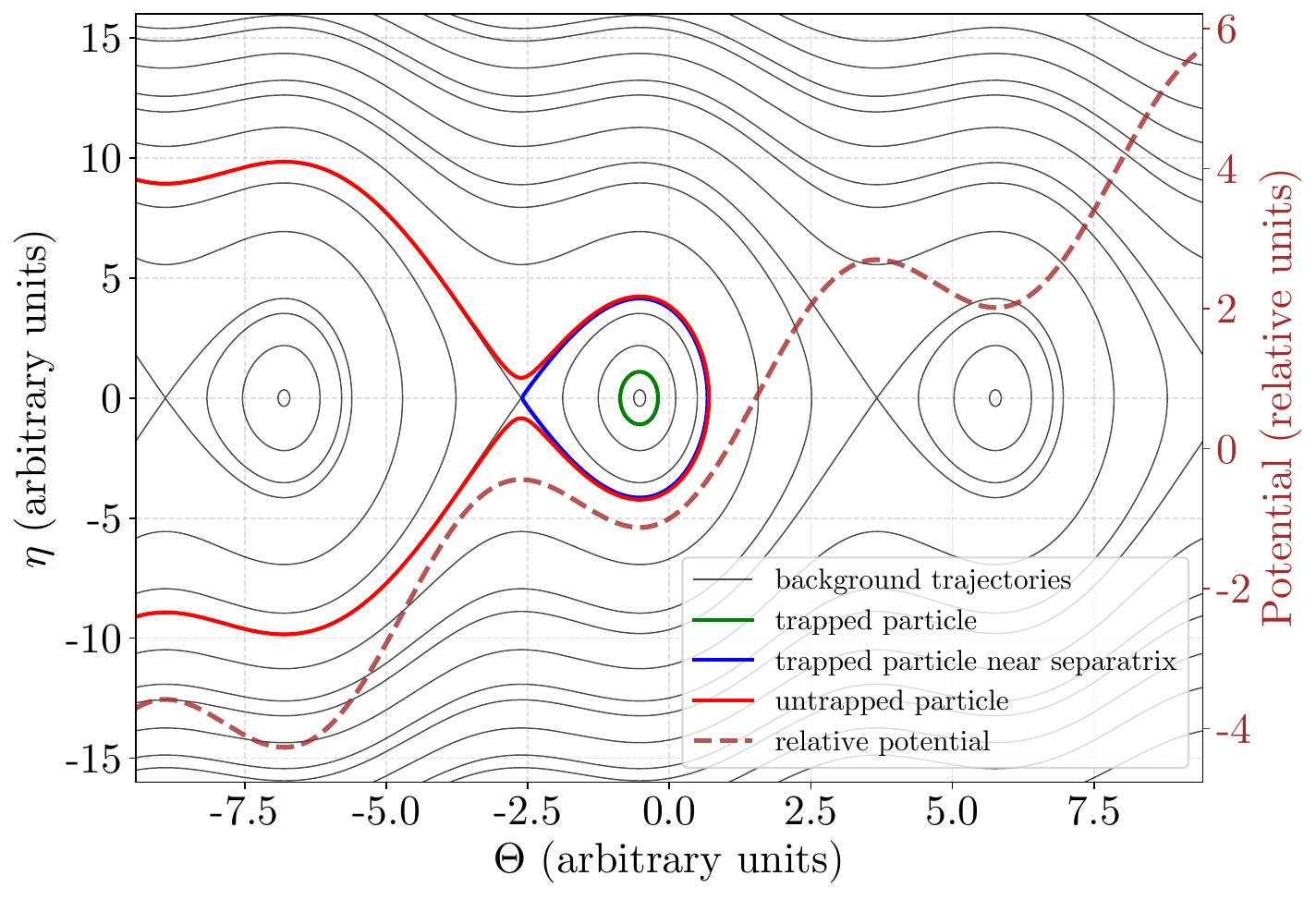}
            \put(10.5,64){\textbf{(a)}}
        \end{overpic}
    \end{subfigure}

    \begin{subfigure}{1\linewidth}
        \centering
        \begin{overpic}[width=\linewidth]{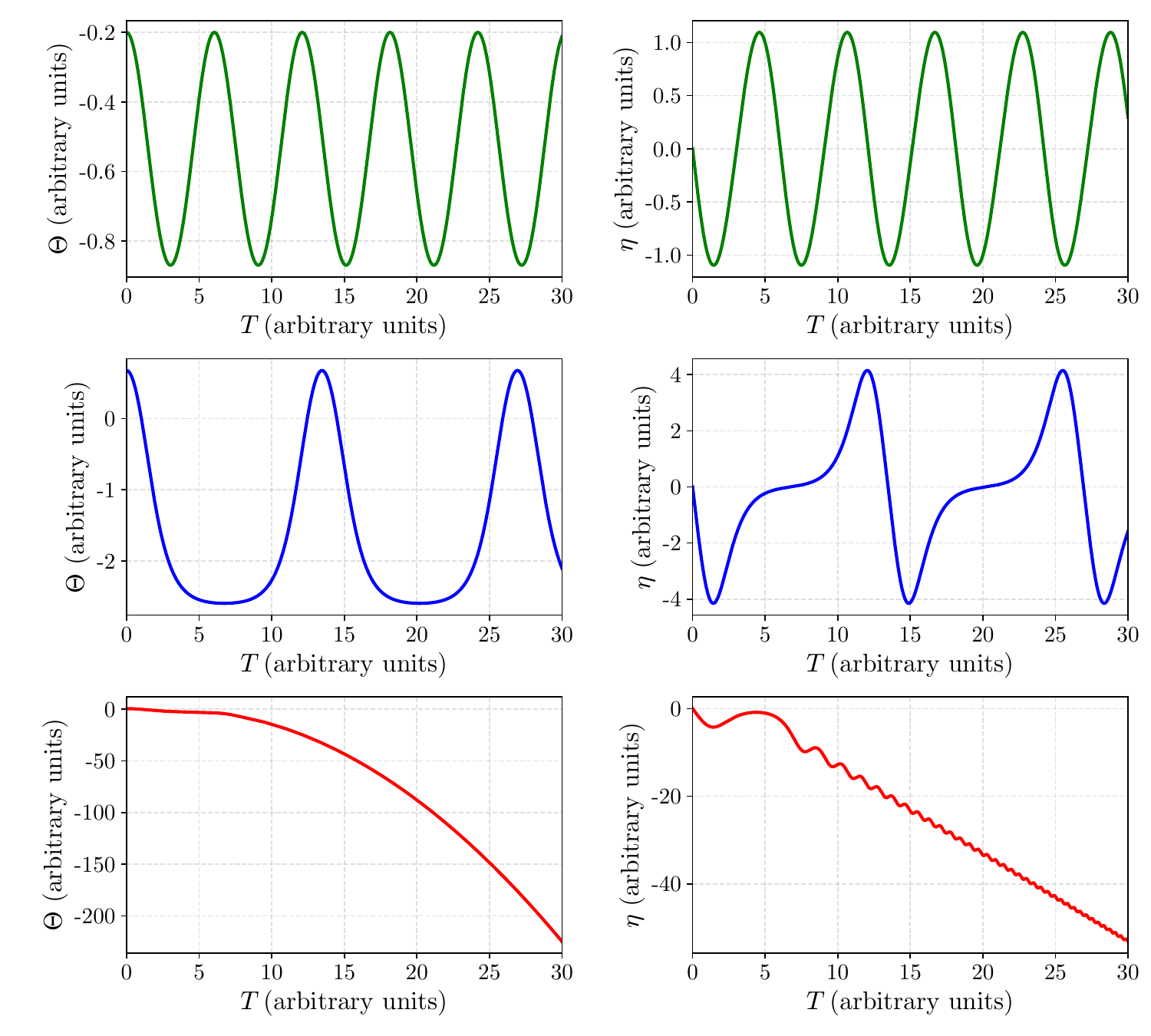}
            \put(12,85){\textbf{(b)}}
            \put(12,55){\textbf{(c)}}
            \put(12,25){\textbf{(d)}}
        \end{overpic}
    \end{subfigure}
    
    \caption{Phase space dynamics in the lattice frame. (a) Velocity phase space trajectories for different particles under lattice force for $\psi=0.5$. Closed contours correspond to trajectories of trapped particles, while open contours represent dynamics of untrapped (perturbed) particles. The three colored curves are drawn for cases of trapped (green), trapped particle near the separatrix (blue), and untrapped (red) particles. The potential variation is shown by the dashed brown curve. Evolution of phase coordinate $\Theta$ and phase velocity $\eta$ against dimensionless parameter $T$ with $\psi=0.5$ for the cases of (b) trapped particle, (c) trapped particle near separatrix, and (d) untrapped particle. Variations are periodic and bounded for both trapped particles. Untrapped particles display monotonic evolution of both dynamical variables accompanied by small lattice induced perturbations.}
\label{fig:phase_space}
\end{figure}

%\av{new text, please check!}
\cref{fig:phase_space} shows the phase space behavior in the accelerated frame of the lattice for an indicative case with $\psi = 0.5$. In \cref{fig:phase_space}(a) the horizontal axis represents the phase coordinate ($\Theta$), the left vertical axis corresponds to the phase velocity ($\eta$), and the right vertical axis shows the potential in relative units. The potential variation (dashed brown curve) and the background trajectories (solid black curves) indicate a series of potential wells associated with the lattice. The closed trajectories represent the dynamics of the trapped particles. Under collisionless conditions, trapped particles indefinitely execute bound oscillatory motion around the stable equilibrium. In the laboratory frame, these trapped particles move together with the lattice. In contrast, the open contours correspond to particles that are not trapped, but nonetheless their motion is affected by the lattice force. In the laboratory frame, these untrapped particles do not move with the lattice, but their velocities can be significantly perturbed by it.

For $\psi=0.5$, one of the potential wells extends from the saddle point $\Theta \approx -2.62$ to $\Theta \approx 0.68$ with a stable equilibrium located around $\Theta \approx -0.52$. Near the well edges, a constant energy trajectory (separatrix) defines the limiting boundary between trapped and untrapped particles. In \cref{fig:phase_space}(a), phase space trajectories are drawn for three representative cases of particles moving around this well. A trapped particle (green) remains bound and orbits around the stable equilibrium point with relatively small phase and phase velocity amplitudes. The particle trapped near the separatrix (blue) also exhibits bound motion with relatively larger amplitudes. An untrapped particle (red) remains unbound, but its motion is significantly influenced: it follows an open trajectory, traversing across neighboring wells without getting trapped.

\cref{fig:phase_space} also presents the dynamics of three particles starting with identical phase velocity ($\eta=0$), but with different initial phases ($\Theta = -0.2,$ $0.675,$ and $0.7$). These initial conditions correspond to the trapped, trapped near the separatrix, and untrapped cases in \cref{fig:phase_space}(a). For both trapped particles, both $\Theta$ and $\eta$ exhibit bounded periodic oscillatory motion within the trapping windows. However, the trapped particle near the separatrix slows down as it reaches the shallow potential barrier to the left (\cref{fig:phase_space}(a)) and spends extended time near $\eta=0$ (\cref{fig:phase_space}(c)).

\cref{fig:phase_space}(d) shows that for the untrapped particle, both $\Theta$ and $\eta$ remain unbounded. The perturbative velocity variations are strongest near $\eta = 0$ and become less and less prominent as $\eta$ diverges away from $0$. Hence, at a given moment in the laboratory frame, an untrapped particle with velocity close to the lattice velocity is strongly perturbed by the lattice. As the velocity mismatch increases, the perturbation effect becomes progressively weaker.

It is noteworthy that a particle starting with a phase velocity within the allowed range for trapping can still escape the lattice. It is thus essential for a particle to fall within the separatrix boundaries to get trapped and move with the lattice. A particle that initially lies outside the separatrix will never be trapped; however, its motion can be significantly influenced by the lattice. For a distribution of particles loaded into the lattice, only a fraction falls within the trapping windows. Hence, the lattice can only trap and effectively accelerate or decelerate a portion of the total population.

\section{Simulation approach}
\label{sec:SimVDF}

Section~\ref{sec:DipForce} discusses the behavior of an individual particle within the lattice. In this section, we discuss a series of numerical simulations to understand how the lattice affects an ensemble of particles. We investigate how the particle velocities change for different interaction cases and identify the conditions for which narrowing of the velocity distribution is possible. To simulate the interaction between a chirped optical lattice and an ensemble of particles, we follow a classical trajectory tracking model similar to the one described in Refs.~\cite{MaherMcWilliams2013_thesis,Gerakis2014_thesis}. The dynamics of all particles are calculated by numerically integrating the equation of motion under the influence of the lattice force.

The electric fields associated with the pulsed laser beams are modeled according to the standard expression for Gaussian beam propagation including the Gaussian intensity envelope, wavefront curvature effect, and Gouy phase. Both laser beams are modeled as having identical spatial beam profiles (TEM$_{00}$), polarization along the same plane, and constant and equal intensities throughout the pulse duration. They differ only in their propagation direction and frequency. The frequency of laser-2 (see \cref{fig:lattice_code}) is varied over the pulse duration to introduce the relative frequency difference $\Delta\omega$ between the two laser beams.

The parameters associated with the simulated laser beams are summarized in \cref{tab:laser_params}; these lie within the capabilities of laser systems used in chirped optical lattice-based experiments~\cite{5th_stage, Bak2022}. The intensity modulation resulting from the interference of two laser beams is estimated numerically and presented in~\cref{fig:lattice_code}. The intensity of a TEM$_{00}$ mode falls with radial distance from the laser axis. We define an effective lattice region within the spatial limits where the intensity of the individual laser beam remains greater than 1/e of its peak value. The interaction is strongest in the effective lattice region outlined by the red diamond geometry with diagonals of $2000$~$\mu$m and $120$~$\mu$m.

\begin{table}[ht]
\caption{\label{tab:laser_params} Laser parameters used in the simulation.}
\begin{ruledtabular}
\begin{tabular}{ll}

Parameter & Value \\

\hline

Laser wavelength $(\lambda)$ & $1064$~nm \\
Beam waist radius $(w_0)$ & $60$~$\mu m$ \\
Half crossing angle $(\theta)$ & $86.75^\circ$ \\
Peak laser intensity $(I_{\max})$ & $1\times10^{13}$~W\,m$^{-2}$ \\
Lattice period $(\lambda_l = \lambda/2\sin\theta)$ & $532.86$~nm \\
Pulse duration $(T)$ & $1000$~ns \\

\end{tabular}
\end{ruledtabular}
\end{table}

\begin{figure}
\centering
\includegraphics[width=0.5\textwidth]{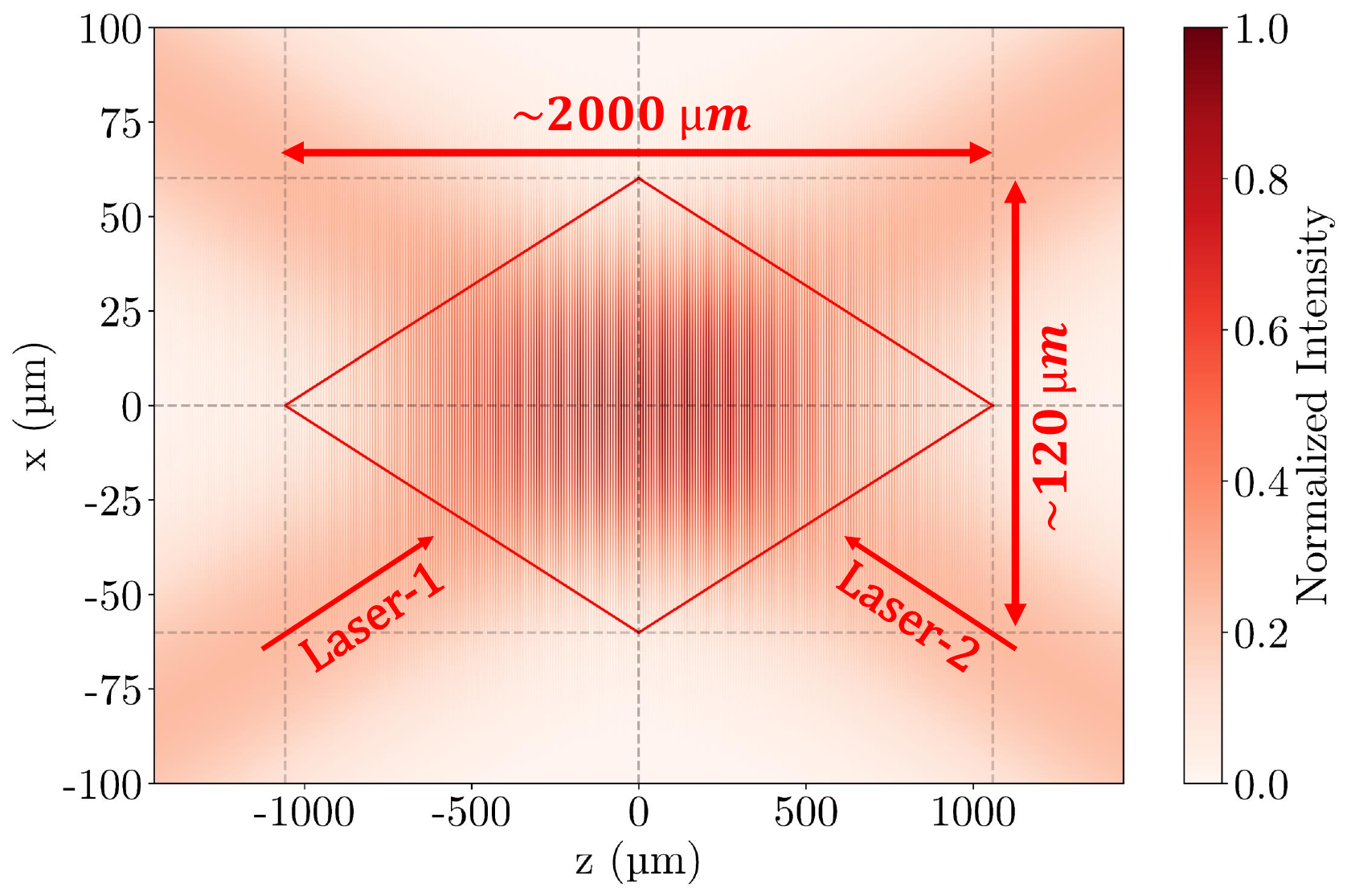}
\caption{Normalized spatial intensity distribution of the lattice formed by crossing of two laser beams. The red diamond shape outlines the effective lattice region.}
\label{fig:lattice_code}
\end{figure}

In \cref{fig:lattice_code} the intensity gradient is significantly steeper along the lattice direction ($z$-axis) than the gradient along the transverse direction ($x$-axis). Consequently, the dipole force in the transverse direction is significantly weaker than the force along the lattice axis. Therefore, the transverse force has negligible influence on particle dynamics compared to the axial force~\cite{MaherMcWilliams2013_thesis, Gerakis2014_thesis}. Hence, the presented study is limited only along the lattice axis (z-axis).

We assume that there are no collisions between the particles in the ensemble and that the dynamics of all particles evolve independently under the effect of the dipole force. Following the definition of dipole force from~\cref{eq:dipole_force}, the equation of motion for a particle having mass $m$ is
\begin{equation}
\frac{d^2z}{dt^2} = -\frac{\alpha_{\text{eff}}}{2m}\frac{\partial}{\partial z}\left[|\mathbf{E_1}(x, y, z, t) + \mathbf{E_2}(x, y, z, t)|^2\right].
\label{eq:eqn_of_motion2}
\end{equation}

%\av{verify with Matthias}
The spatial derivative is evaluated at each time step during numerical integration using automatic differentiation (via PyTorch's autograd package~\cite{paszke2019pytorch}). This allows for convenient, numerically stable, and accurate simultaneous computation of particle accelerations. The particle trajectories are tracked by integrating \cref{eq:eqn_of_motion2} using the velocity-Verlet algorithm~\cite{Verlet_algorithm} with a time step of $0.1$~ns. The validity of this value was validated through convergence tests. The simulation is fully vectorized over all particles, facilitating simultaneous evaluation of all particles.

%distribution 

We simulate the interaction of the lattice with a representative test-case ensemble of $10^5$ neutral cesium atoms (atomic mass = $132.9$~amu). The ground-state cesium atom exhibits a large static polarizability $\alpha_{\text{eff}}=6.611\times10^{-39}$~Cm$^2$V$^{-1}$ ~\cite{Polarizability}. The laser wavelength of $1064$~nm is far off-resonance to drive transitions in a ground state cesium atom, ensuring it remains in the ground state. The laser intensities of the order of $10^{13}$~W$\cdot$m$^{-2}$ result in a dipole potential well depth of $1.8$~K ($\approx 0.16$~meV). All atoms are uniformly distributed over a cylindrical volume of radius $30$~$\mu$m and length $250$~$\mu$m. The initial velocities are randomly sampled following a Gaussian distribution with a mean at $1000$~ms$^{-1}$, a standard deviation of $100$~ms$^{-1}$ and truncated to lie in between $700$~ms$^{-1}$ and $1300$~ms$^{-1}$. The initial positions and initial velocities of the particles are assumed to be non-correlated (see \cref{fig:Initial_distribution}).

\begin{figure}
\centering
\includegraphics[width=0.5\textwidth]{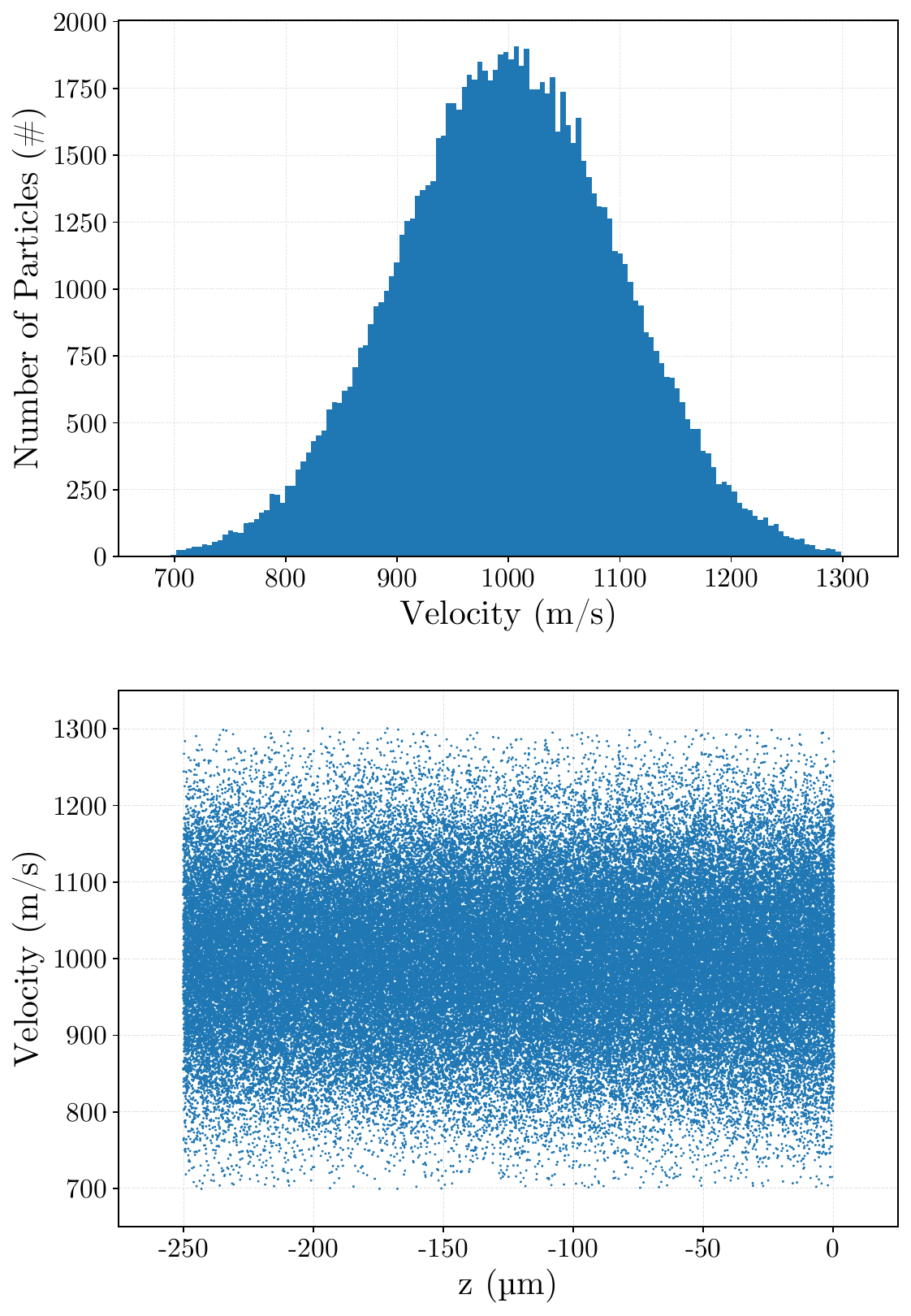}
    \put(-220,350){\textbf{(a)}}
    \put(-220,160){\textbf{(b)}}
\caption{Initial distribution of cesium atoms. (a) Initial velocity distribution is sampled from a Gaussian centered at $1000$~ms$^{-1}$, with standard deviation $100$~ms$^{-1}$, and limited from $700$~ms$^{-1}$ to $1300$~ms$^{-1}$. (b) Initial phase space distribution showing; atoms are uniformly distributed between $-250$~$\mu$m and $0$~$\mu$m.}
\label{fig:Initial_distribution}
\end{figure}

The hard sphere collision estimation for the defined test ensemble with van der Waals radius $343$~pm~\cite{van_der_Waals_Radi} corresponds to a mean free path of $\approx42.8$~m and a mean collision time of $\approx42.8$~ms. The estimated mean free path is many orders ($\approx2\times10^{4}$) larger than the length of the effective lattice region, and the mean collision time is much longer ($\approx4\times10^{4}$ times) than the laser pulse duration. Therefore, under these conditions, collisions between cesium atoms can be safely neglected.

\section{Interaction cases and velocity distribution narrowing}
\label{sec:Sim_cases}

To understand how the chirped optical lattice affects the ensemble defined in \cref{sec:SimVDF}, we simulate and analyze different interaction cases with varying chirp schemes and interaction approaches. We follow the redistribution of cesium atoms over the velocity space and identify conditions for which the optical lattice can narrow the velocity distribution. Finally, we design and simulate an interaction approach that demonstrates the desired narrowing of the velocity distribution.

First, the ensemble defined in \cref{fig:Initial_distribution} is loaded in the optical lattice (\cref{fig:lattice_code}). The complete ensemble remains within the boundaries of the optical lattice. Hence, all cesium atoms are affected by the dipole force simultaneously. The lattice velocity is swept from $1000$~ms$^{-1}$ to $1300$~ms$^{-1}$ linearly (see \cref{fig:extraction} (a)) over the duration of the laser pulse of $1000$~ns (interaction case-A).

The redistribution of cesium velocities for interaction case-A is shown in \cref{fig:extraction}. A portion of the population is accelerated to higher velocities, resulting in peak formation near the velocity $1300$~ms$^{-1}$ (see \cref{fig:extraction}(b)). To understand the redistribution of the cesium population, the difference of the final and initial velocity is plotted against the initial velocity for each atom in \Cref{fig:extraction}(c). The lattice traps and accelerates cesium atoms starting near the initial velocity $1000$~ms$^{-1}$. A portion of these atoms is efficiently trapped and accelerated to final velocities close to $1300$~ms$^{-1}$. This corresponds to the formation of the peak in \cref{fig:extraction}(b). The remaining trapped atoms are accelerated to lower final velocities, resulting in population growth in the range $1150$~ms$^{-1}$ to $1300$~ms$^{-1}$ (\cref{fig:extraction}(b)). These atoms start with initial positions and velocities such that they move out of the high intensity central lattice region during the interaction. Hence, they drop out before the lattice finishes the velocity sweep and undergo only partial acceleration.

In addition to trapping and acceleration, the lattice also perturbs a significant population with initial velocities within the range of the lattice velocity sweep ($1000$~ms$^{-1}$ to $1300$~ms$^{-1}$). This perturbation appears as particles dispersed away from the $\Delta v = 0$ line in the velocity sweep range. Particles outside the range (initial velocity < $1000$~ms$^{-1}$) remain largely unaffected or experience only weak perturbations. Although the lattice transfers the population from the center to the right periphery of the velocity distribution, a sharp depletion of the population in the central region of the velocity distribution is not observed. This is due to the perturbative deceleration that replenishes the population near the mean velocity.

Despite the perturbations near the right periphery, the net population in this region increases. A low initial population near the right periphery results in a relatively lower number of atoms being perturbed. Hence, the population transfer via trapping and acceleration from the center to the right periphery is more prominent than the perturbations.

\begin{figure}
\centering
\includegraphics[width=0.5\textwidth]{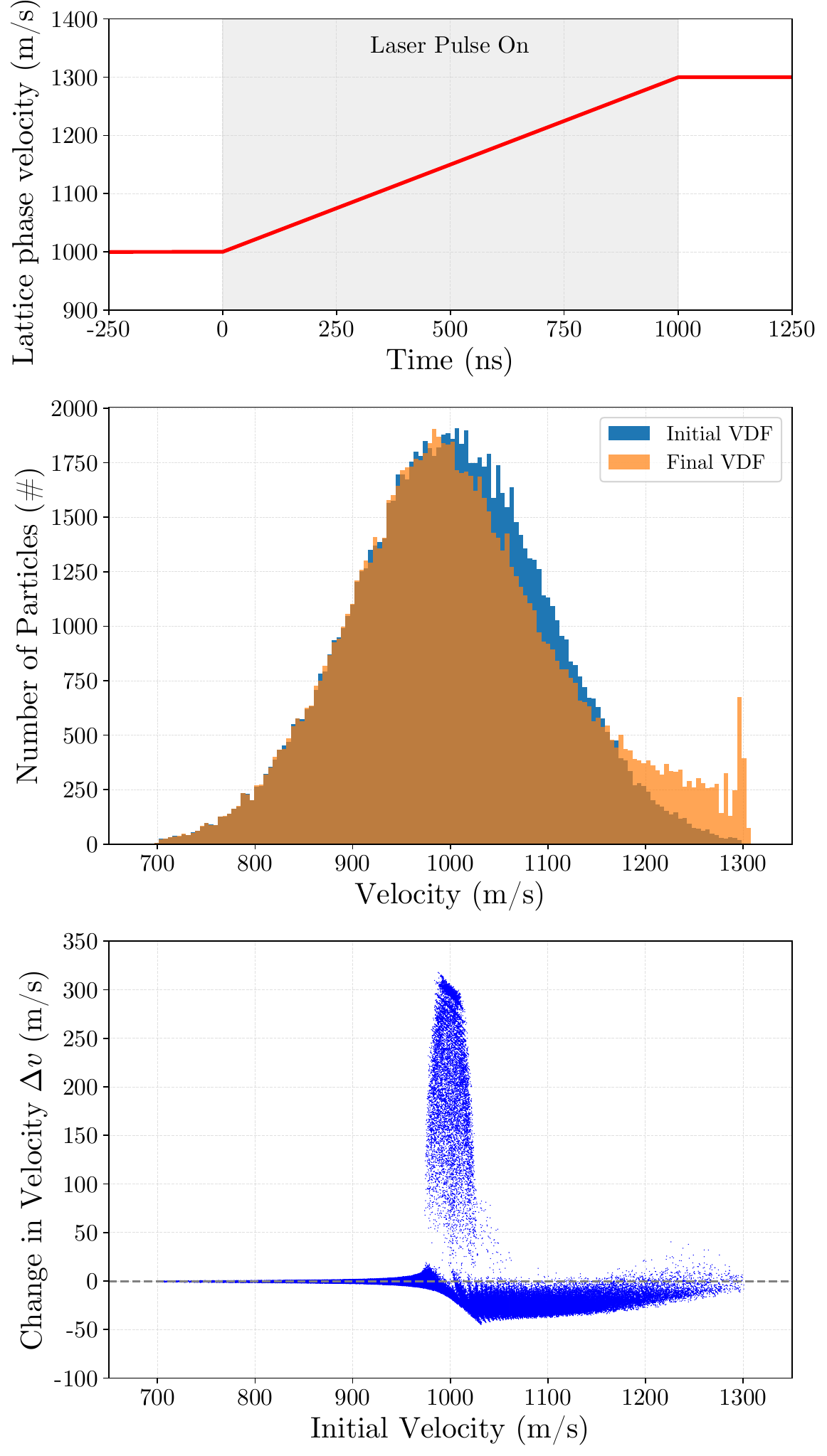}
    \put(-220,433){\textbf{(a)}}
    \put(-220,310){\textbf{(b)}}
    \put(-220,148){\textbf{(c)}}
\caption{Interaction case-A. (a) Variation in lattice phase velocity against time. Negative time used only for the representation. t=0 marks the instance when laser pulse is turned on. (b) Comparison of initial (blue) and final (orange) velocity distributions. The lattice transfers a population portion towards the right periphery. (C) Change in velocity vs initial velocity for all atoms. A portion of particles starting near $1000$~ms$^{-1}$ is trapped and accelerated to higher velocities. The lattice also perturbs a significant part of the population.}
\label{fig:extraction}
\end{figure}

To effectively narrow the velocity distribution, the lattice must transfer cesium atoms from the periphery to the central region of the velocity distribution. To explore this possibility, we study the interaction case (case-B) with an identical phase space configuration of the ensemble as used in interaction case-A (\cref{fig:Initial_distribution}). The lattice is made to decelerate linearly from an initial velocity of $1300$~ms$^{-1}$ to a final velocity of $1000$~ms$^{-1}$ over the duration of $1000$~ns (see \cref{fig:dip} (a)).

However, this interaction case does not result in the accumulation of new atoms near the mean velocity. The final velocity distribution exhibits population depletion near the central region (see \cref{fig:dip}(b)). In addition, the distribution broadens slightly near higher velocities, indicating small acceleration of the particles. \cref{fig:dip}(c) shows that only a very small fraction of the cesium population starting near $1300$~ms$^{-1}$ is trapped and decelerated towards lower velocities. Some of the trapped particles are partially decelerated because they move out of the effective lattice region before the lattice velocity sweep finishes. Particles in the lattice velocity sweep range experience perturbative acceleration, which causes a slight broadening of the velocity distribution.

Initially, the lattice is turned on with a velocity of $1300$~ms$^{-1}$ where the density of the velocity distribution is lower. Few high velocity atoms get trapped and are decelerated with the lattice. The lattice also continues to perturb the population encountered during the linear deceleration. As the lattice velocity approaches near $1000$~ms$^{-1}$, it perturbs a significantly larger number of cesium atoms. The number of particles decelerated to the final velocity $1000$~ms$^{-1}$ is much lower than the number of cesium atoms perturbed away from region near the velocity $1000$~ms$^{-1}$, resulting in a net population depletion at the mean velocity.

From \cref{fig:dip}(c) it can be seen that the lattice transfers some population from the periphery to the center of the velocity distribution. The depletion near the mean velocity is caused by lattice induced velocity perturbations. Hence, to narrow the velocity distribution, the lattice must transfer atoms from the periphery to the center of the velocity distribution without perturbing the existing population near the center.

Such selective manipulation can be made possible by introducing a free space propagation of the original ensemble before it interacts with the lattice. The ensemble stretches along the propagation direction (z-axis) since it contains particles with different velocities. Faster particles move towards the front and slower particles follow, establishing a correlation between particle velocities and positions. As the elongated ensemble is guided towards the effective lattice interaction region, the interaction can be timed in a way such that only fast atoms interact with the lattice. The perturbative depletion of the population near mean velocity can thus be minimized by precisely timing the interaction.

To ensure that the lattice does not affect particles having velocities near the mean velocity, the ensemble defined in \cref{fig:Initial_distribution}(b) is allowed to propagate in free space until its total length reaches approximately $3000$~$\mu$m. The expansion takes place over a distance of about $5.9$~mm and an approximate duration of $4.6$~$\mu$s. The optical lattice is turned on immediately after this free-space propagation (see \cref{fig:push}). It decelerates linearly from the initial velocity of $1300$~ms$^{-1}$ to the final velocity $1000$~ms$^{-1}$ over the pulse duration of $1000$~ns (\cref{fig:push}(a)). To maximize the population transferred from the periphery to the center, different spatial placements of the initial packet with respect to the origin (z=0) were tested. The most efficient population transfer was observed for the initial packet placed in the approximate spatial range $-6.85$~mm to $-6.6$~mm (see \cref{fig:drift}(a)). The phase space distribution after the free-space propagation is presented in \cref{fig:drift}(b). The fast atoms lead the packet and the slower atoms lag behind. At the end of the free-space propagation, the packet front is partially inserted in the effective lattice region.

\begin{figure}
\centering
\includegraphics[width=0.5\textwidth]{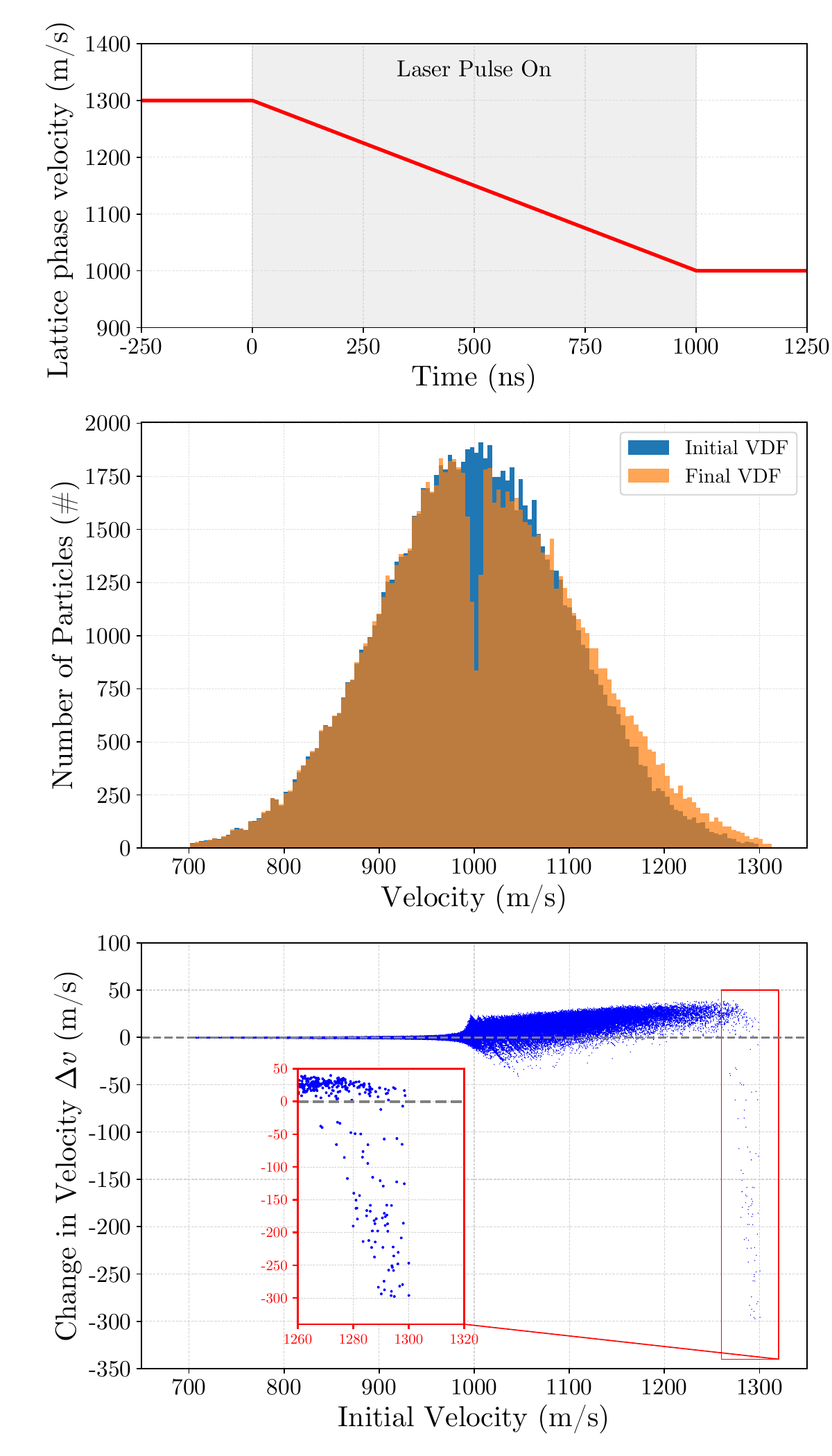}
    \put(-210,422){\textbf{(a)}}
    \put(-210,305){\textbf{(b)}}
    \put(-210,147){\textbf{(c)}}
\caption{Interaction case-B. (a) Variation in lattice phase velocity against time. Negative time used only for the representation. t=0 marks the instance when laser pulse is turned on. (b) Comparison of initial (blue) and final (orange) velocity distributions. The lattice depletes population near the mean velocity. (c) Change in velocity vs initial velocity for all atoms. A very small fraction of population near the periphery is trapped and decelerated toward mean velocity. Near velocity $1000$~ms$^{-1}$ the perturbation depletes more population than the lattice brings in via trapping and deceleration. The inset shows the variations near high initial velocity region with adjusted axis limits, tick scaling and scatter point size for visual clarity.}
\label{fig:dip}
\end{figure}

With this interaction scheme (case-C), the particle population increases significantly near the mean velocity as seen in \cref{fig:push}(b). Atoms with initial velocities of more than $1000$~ms$^{-1}$ decelerate and move towards the center of the velocity distribution. \cref{fig:push}(c) suggests that atoms through the lattice velocity sweep range are trapped and decelerated near a velocity of $1000$~ms$^{-1}$. It is also observed that atoms near the initial of velocity $1000$~ms$^{-1}$ are slightly perturbed. However, the perturbation effect is less prominent than trapping and deceleration, and a net population gain is observed near $1000$~ms$^{-1}$.

\begin{figure}
\centering
\includegraphics[width=0.5\textwidth]{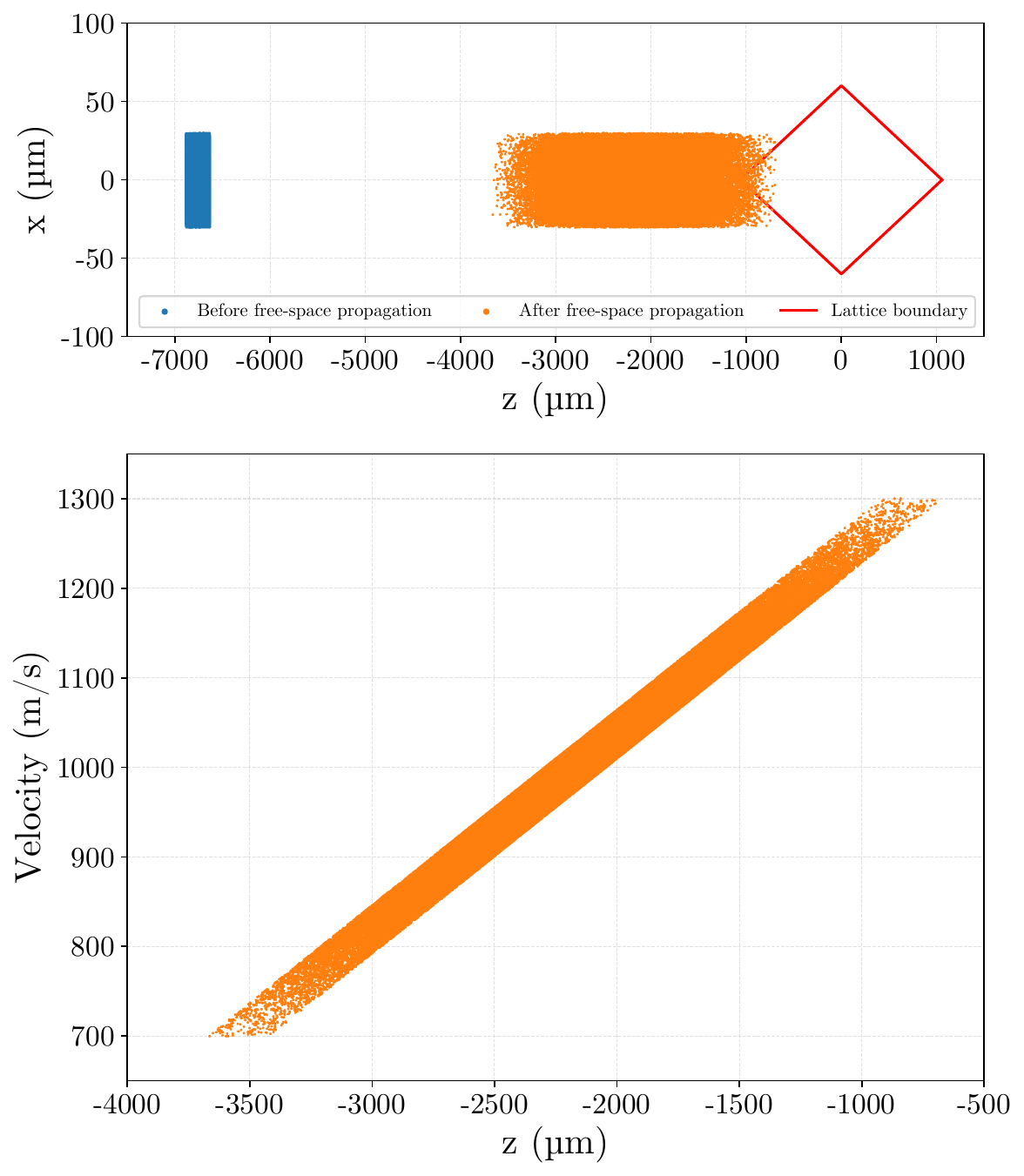}
    \put(-220,275){\textbf{(a)}}
    \put(-220,165){\textbf{(b)}}
\caption{Effect of free-space propagation on the ensemble. (a) The free-space propagation stretches the ensemble along the $z$-axis. (b) Phase space distribution after the free-space propagation. Faster atoms lead the packet and slower atoms follow. The velocity and positions become correlated.}
\label{fig:drift}
\end{figure}

\begin{figure}
\centering
\includegraphics[width=0.5\textwidth]{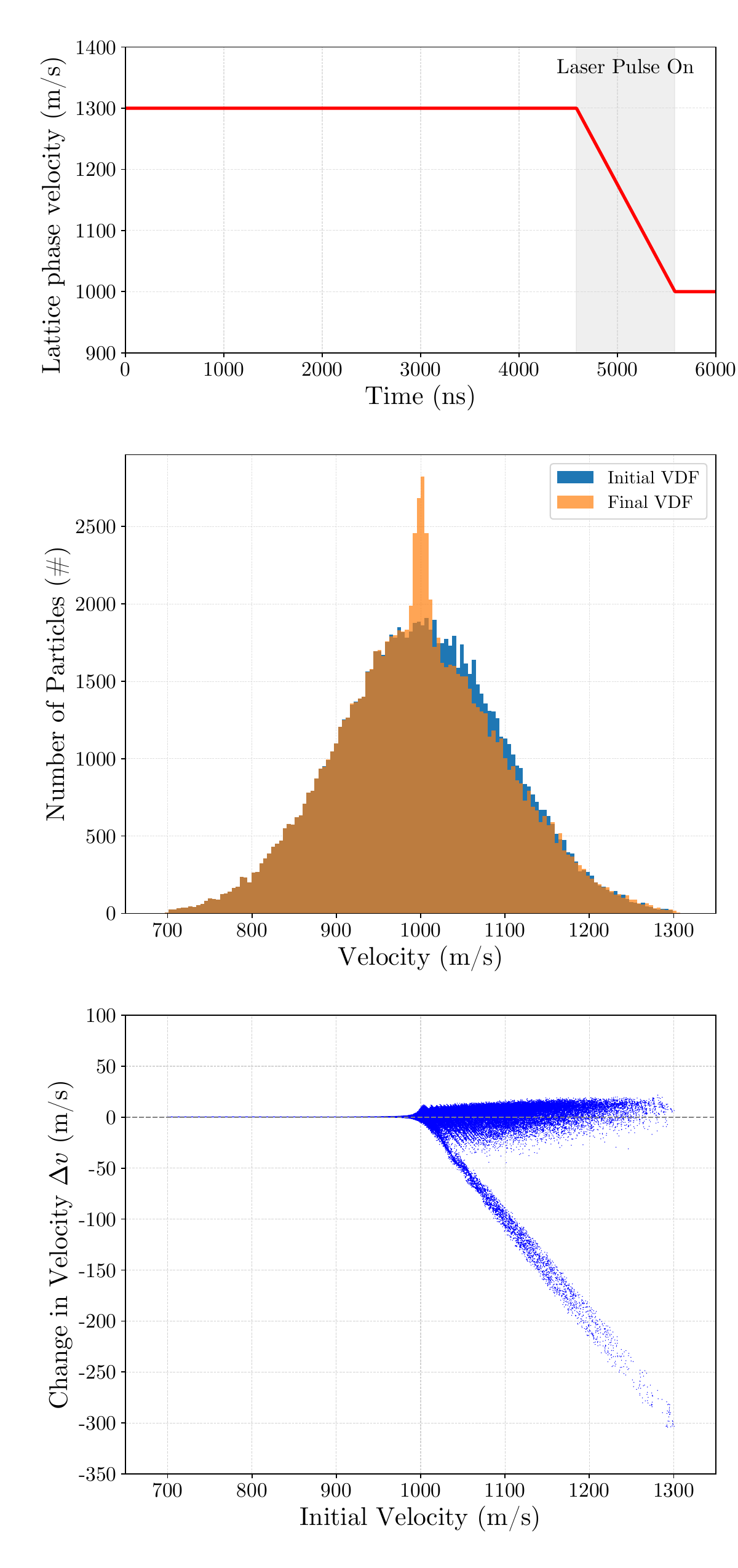}
    \put(-210,515){\textbf{(a)}}
    \put(-210,370){\textbf{(b)}}
    \put(-210,180){\textbf{(c)}}
\caption{Interaction case-C. (a) Variation in lattice phase velocity against time. (b) Comparison of initial (blue) and final (orange) velocity distributions. The lattice increases significant population near velocity $1000$~ms$^{-1}$. (c) Change in velocity vs initial velocity for all atoms. Atoms in the complete range of lattice velocity sweep are trapped and decelerated.}
\label{fig:push}
\end{figure}

In the first two interaction cases, the lattice only traps atoms within a small velocity window and accelerates or decelerates them (see \cref{fig:extraction}(c) and \cref{fig:dip}(c) respectively). In contrast, in the interaction case-C, trapping and deceleration occur over the complete range of the lattice velocity sweep (\cref{fig:push}(c)). In the first two cases, the initial phase space distribution is compactly localized along the $z$-axis and positioned entirely within the effective lattice region. These initial conditions allow the atoms to move out of the effective lattice region before the lattice completes the velocity sweep. In the free-space propagation assisted case, the packet enters the effective lattice region as the lattice slows down. At a given moment, if the velocity of an atom entering the effective lattice region is close to the lattice velocity, it can be trapped and decelerated with the lattice. This results in trapping and deceleration of atoms in the complete lattice velocity sweep range.

%%\av{do we need a concluding paragraph here? I am continuing the discussion in the next section.}

\section{Discussion and Conclusion}
The interaction cases discussed in \cref{sec:Sim_cases} demonstrate that a chirped optical lattice can not only extract a population packet from a velocity distribution as reported in previous studies ~\cite{Barker2001, Barker2002, Gerakis2014_thesis, Maher-McWilliams2012_paper, MaherMcWilliams2013_thesis}, but it can also narrow the velocity distribution under appropriate interaction conditions. The enabling mechanism for narrowing is to selectively interact with and decelerate only the fast particles while keeping the remaining population unaffected.

The pre-interaction free-space propagation stage facilitates such selective manipulation of particle velocities by restructuring the phase space distribution. The ensemble stretches along the propagation direction, developing a velocity-position correlation. The interaction is timed so as to effectively manipulate only the fast particles of the ensemble. Particles near the center of the distribution do not experience strong perturbations, while particles from high velocity regions are trapped and decelerated. The net effect is the accumulation of the population near the center of the velocity distribution.

%change, we fixed the length of the extended ensemble and optimized the offset according to that.

The initial ensemble is allowed to stretch to a length of $~3000~\mu m$ prior to the lattice interaction. Subsequently, to maximize population transfer, we scan the position offsets for the initial ensemble with respect to the origin, while keeping the interaction time constant ($1000$~$ns$). By controlling the offset, we tune the phase space region that is exposed to the effective lattice region and identify the configuration for which the net population accumulation is maximal near velocity $1000$~ms$^{-1}$. Similar scans can also be performed for different expansion lengths and pulse durations to further optimize population transfer in the desired direction.

In this study, the acceleration of the lattice and the laser intensity remain constant over the laser pulse duration. The technique could be improved further through introduction of nonlinear lattice velocity schedules and time-dependent laser intensity variations. These parameters dictate the capture windows and efficiency, as discussed in Section-II. Time-dependent lattice acceleration and laser intensities could be utilized to maximize the trapping and minimize the perturbations near the final target velocity.

In the working interaction case-C, the lattice transfers the population from the high-velocity periphery to the center of the velocity distribution by decelerating fast particles. In principle, the population could also be transferred from the low-velocity periphery to the center. This would require a more elaborate interaction design due to the asymmetry between the acceleration and deceleration processes. The faster particles of the elongated ensembles will enter the lattice region first, followed by the slower particles. However, an accelerating lattice will start the velocity sweep from slow to fast velocity. This mismatch of the arrival sequence of the particles and the lattice velocity evolution would limit the efficiency of trapping and acceleration. The development of such technique coupled with two or more timed optical lattices can compress the velocity distribution of a particle distribution via both acceleration and deceleration over the same cycle.

The approach presented here relies on free-space propagation to introduce spatial velocity separation. For this, the free-space propagation length should be sufficient to stretch the ensemble appropriately. Also, the laser should be timed so that the lattice does not interact with particles close to the mean velocity. In addition, the trapping velocity window \(\Delta v = 2 \sqrt{2\Delta U/m}\) depends on the laser intensity and the chirp rate and should always be lower than the width of the final desired velocity distribution. Otherwise, the velocity spread of trapped particles itself will be greater than the final velocity spread desired. In addition, very high laser intensities can lead to multiphoton ionization~\cite{Tropina02062016, mnhl2005}. Finally, since the electric dipole force is purely conservative by nature, it cannot compress the phase space volume according to Liouville's theorem~\cite{Reinaudi_2008}. Hence, the approach redistributes the particles over the phase space retaining the phase space volume instead of compressing the distribution. This fundamentally distinguishes the technique presented here from dissipative laser cooling.

Apart from cesium atoms, the method is in principle applicable to other polarizable particles (atoms, molecules, and ions) given the lasers remains far off-resonance. The extension of the presented approach to ion ensembles would be particularly relevant for instruments designed to focus ion beams onto small areas (probe size) for numerous applications~\cite{fit4nano}. In such instruments, ions with different kinetic energies are focused at different axial positions, limiting the minimum achievable probe size. This effect is dubbed chromatic aberration~\cite{Hawkes2009} and can be minimized by reducing the velocity spread of an ion beam. However, for charged particle ensembles, the ion densities must remain low enough so that space-charge effects do not dominate over the lattice dipole force.

%restructure here1
In summary, we have demonstrated that a chirped optical lattice, when combined with inducing a position–velocity correlation via free-space propagation, can be used to narrow the particle velocity distribution. By restricting the interaction to a spatially separated subset of the ensemble, the competing perturbative effects that limit conventional schemes were avoided, enabling the controlled transport of particles to the center of the velocity distribution. This establishes a new operational regime for chirped optical lattices and provides a non-resonant approach to velocity-space manipulation with broad applicability.

\section{Acknowledgments}
All authors are supported by the Luxembourg National Research Fund 17382436 (VERITAS). The authors acknowledge Dr. Stefan Karatodorov, Dr. Maria Mitrou, Dr. Marios Kounalakis, and Dr. Ali Hosseinnia for assistance with the manuscript.

\bibliographystyle{apsrev4-2}
\bibliography{references}

\end{document}